\documentclass[aps,preprint,superscriptaddress]{revtex4-2}

\usepackage[hidelinks]{hyperref}
\usepackage{amsmath, graphicx}

\begin{document}

\begin{flushright}
{
OU-HET-1223
}
\end{flushright}

\title{
Exploring loop-induced first-order electroweak phase transition\\
in the Higgs effective field theory 
}

\author{Ricardo R. Florentino}
\email{rflorentino@het.phys.sci.osaka-u.ac.jp}
\affiliation{
Department of Physics, Osaka University, Toyonaka, Osaka 560-0043, Japan
}

\author{Shinya Kanemura}
\email{kanemu@het.phys.sci.osaka-u.ac.jp} 
\affiliation{
Department of Physics, Osaka University, Toyonaka, Osaka 560-0043, Japan
}

\author{Masanori Tanaka}
\email{tanaka@pku.edu.cn}
\affiliation{
Center for High Energy Physics, Peking University, Beijing 100871, China
}

\begin{abstract}

The nearly aligned Higgs Effective Field Theory (naHEFT) is based on the general assumption: all deviations in the Higgs boson couplings are originated from quantum one-loop effects of new particles that are integrated out.
If the new particles integrated out have the same non-decoupling property, physics of the electroweak symmetry breaking can be then described by several parameters in the naHEFT, so that there is a correlation among the Higgs boson couplings such as $h \gamma \gamma$, $hWW$ and $hhh$ couplings. 
In this paper, we analyze the strongly first-order electroweak phase transition (EWPT) with the condition of sphaleron decoupling and the completion condition of the phase transition, and investigate the relation among the deviations in the Higgs boson couplings and the dynamics of the EWPTs. 
We also take into account the gravitational wave spectrum as well as the primordial black hole predicted at the EWPT.
We show that if the new particles integrated out include charged scalar states future precision measurements of the $h \gamma \gamma$ coupling can give a useful prediction on the $hhh$ coupling to realize the strongly first-order EWPT. 
We can explore the nature of EWPT and the new physics behind it by the combination of precision measurements of various Higgs boson couplings at future collider experiments, gravitational wave observations at future space-based interferometers and searches for primordial black holes. 

\end{abstract}

\maketitle
\newpage

%%%%%%%%%%%%%%%%%%%%%%%%%%%%%%%%%%%%%%%%%%%%%%

\section{Introduction}

The standard model (SM) of particle physics is successful in explaining the current results at collider experiments. 
However, the SM cannot explain several phenomena such as the baryon asymmetry of the Universe (BAU)~\cite{Planck:2018vyg}, the existence of dark matter~\cite{Planck:2018vyg} and neutrino oscillation~\cite{Super-Kamiokande:1998kpq}.
In addition, the structure of the Higgs sector, which is important to realize the spontaneous symmetry breaking, is still unknown.

The electroweak baryogenesis (EWBG) is a promising scenario, in which the baryon asymmetry can be produced by the dynamics of electroweak symmetry breaking~\cite{Kuzmin:1985mm}. 
The EWBG requires the electroweak phase transition (EWPT) to be first-order in order to satisfy the third Sakharov's condition~\cite{Sakharov:1967dj}.
To realize the EWBG scenario, the sphaleron process should decouple immediately inside symmetry broken regions. 
This condition is often called the sphaleron decoupling condition, which is approximately expressed by~\cite{Kuzmin:1985mm} 
\begin{align}
\frac{v_{n}}{T_{n}} > 1, 
\label{eq:vnTn}
\end{align}
where $T_{n}$ is the nucleation temperature when a vacuum bubble is nucleated in a Hubble horizon, and $v_{n}$ is the vacuum expectation value at $T_{n}$. 
Although the right-hand side of the sphaleron decoupling condition is often taken as unity, it mildly depends on the sphaleron energy in new physics models~\cite{Ahriche:2007jp,Funakubo:2009eg,Fuyuto:2014yia,Ahriche:2014jna,Fuyuto:2015jha, Kanemura:2020yyr, Kanemura:2022ozv}. 

In the SM, the EWBG cannot be realized because the EWPT is not first-order~\cite{Kajantie:1996mn,DOnofrio:2014rug,DOnofrio:2015gop}. 
In addition, the CP violation in Cabbibo-Kobayashi-Maskawa phase is too small to explain the observed baryon asymmetry~\cite{Gavela:1994dt,Huet:1994jb}. 
Therefore, an extension of the Higgs sector is needed to realize the EWBG. 
The EWBG in various extended Higgs models has been discussed~\cite{Turok:1990in,Turok:1990zg,Cline:1995dg,Carena:1997gx,Cline:1998hy,Cline:1997vk,Cline:2000nw,Cline:2000kb,Carena:2000id,Kainulainen:2002th,Carena:2002ss,Cirigliano:2006wh,Fromme:2006cm,Chung:2009qs,Li:2008ez,Cline:2011mm,Liu:2011jh,Cline:2012hg,Alanne:2016wtx,Chiang:2016vgf,Inoue:2015pza,Jiang:2015cwa,Cline:2017qpe,Guo:2016ixx,Vaskonen:2016yiu,Fuyuto:2017ewj,Grzadkowski:2018nbc,Ramsey-Musolf:2017tgh,Modak:2018csw,Fuyuto:2019svr,Cline:2021iff,Kainulainen:2021oqs,Xie:2020wzn,Enomoto:2022rrl,Enomoto:2021dkl,Aoki:2023xnn,Aoki:2022bkg,Basler:2021kgq,Kanemura:2023juv}.

It is expected that significant deviations in Higgs couplings from the SM prediction are required to satisfy the sphaleron decoupling condition in Eq.\,\eqref{eq:vnTn} in new physics models.
The sphaleron decoupling condition requires large deviations in the triple Higgs boson coupling $hhh$, where $h$ is the SM Higgs field~\cite{Grojean:2004xa, Kanemura:2004ch, Kakizaki:2015wua, Hashino:2016rvx}.
For instance, in the two Higgs doublet model, the sphaleron decoupling condition predicts more than about $+30\,\%$ deviations in the $hhh$ coupling~\cite{Enomoto:2021dkl, Enomoto:2022rrl, Kanemura:2022ozv}. 
Therefore, the nature of the EWPT can be explored by the precise $hhh$ coupling measurements. 
The deviation in the $hhh$ coupling can be realized by scalar mixing effects or radiative corrections~\cite{Kanemura:2004mg}.
In general, the scalar mixing effects at tree level make the $hhh$ coupling smaller than the SM prediction.
On the other hand, the $hhh$ coupling with bosonic radiative corrections is larger than the SM one.
The scalar mixing effects are strongly constrained by current Higgs coupling measurements~\cite{ATLAS:2022vkf,CMS:2022dwd}. 
Therefore, the radiative corrections can make the deviation in the $hhh$ coupling large.
Although the $hhh$ coupling has not been measured precisely so far~\cite{CMS:2022dwd,ATLAS:2022jtk}, 
it may be measured with higher precision accuracy at future collider experiments.
For instance, it is expected that the $hhh$ coupling may be measured with about 50\,\% and 10\,\% accuracies at the High-Luminosity LHC (HL-LHC)~\cite{Cepeda:2019klc} and the International Linear Collider (ILC) with $\sqrt{s} = 1\,{\rm TeV}$~\cite{Bambade:2019fyw}, respectively.  

We note that Higgs couplings other than the $hhh$ coupling are also important to test the nature of the EWPT. 
The radiative corrections from new particles, which make the $hhh$ coupling large, can realize large deviations in Higgs couplings to gauge bosons such as $hZZ$~\cite{Kanemura:2004mg} and $h\gamma \gamma$~\cite{Shifman:1979eb}. 
In particular, as the $h \gamma \gamma$ coupling is induced by loop effects, it is deeply related to the properties of new charged particles.
The deviations in the Higgs couplings have been discussed in various extended Higgs models like two Higgs doublet models~\cite{Arhrib:2003vip,Aoki:2009ha,Posch:2010hx,Arhrib:2012ia,Chiang:2012qz,Fontes:2014xva,Kanemura:2014dja,Kanemura:2014bqa,Arhrib:2015hoa,Kanemura:2015mxa,Hashino:2015nxa,Kanemura:2016sos,Senaha:2018xek,Braathen:2020vwo,Florentino:2021ybj,Degrassi:2023eii,Aiko:2023nqj,Aiko:2023xui}, Higgs triplet models~\cite{Akeroyd:2010je,Arhrib:2011vc,Aoki:2012yt,Kanemura:2012rs,Aoki:2012jj,Arbabifar:2012bd,Chiang:2012qz}, the SM with singlet scalar fields~\cite{Shifman:1979eb, Chiang:2012qz,Katz:2014bha,Kakizaki:2015wua,Kanemura:2015fra,Kanemura:2016lkz,Braathen:2019pxr,Braathen:2019zoh,Aiko:2023xui}, and the Georgi-Machacek model~\cite{Georgi:1985nv, Chiang:2017vvo,Chiang:2018xpl}.
The LHC has measured the Higgs couplings to weak gauge bosons $hVV\,(V = W, \, Z)$ and fermions $hff$ with about $10\,\%$ accuracy~\cite{ATLAS:2022vkf,CMS:2022dwd}. 
The loop-induced Higgs couplings $h \gamma \gamma$ and $h Z \gamma$ have been also measured at the LHC with about $11\,\%$ and $40\,\%$ accuracies, respectively~\cite{ATLAS:2022vkf, ATLAS:2022tnm, CMS:2022dwd}. 
The $hVV$, $hff$ and $h \gamma \gamma$ couplings can be measured with more precise accuracy at future collider experiments such as the HL-LHC~\cite{Cepeda:2019klc}, ILC~\cite{Bambade:2019fyw}, CLIC~\cite{Roloff:2018dqu} and FCCs~\cite{FCC:2018byv}.
Especially, it is expected that the $h \gamma \gamma$ coupling can be measured with $1.8\,\%$ accuracy ($68\,\%$ confidence level) at the HL-LHC~\cite{Cepeda:2019klc}.

In addition to the Higgs coupling measurements, direct new physics searches are also quite important. 
For instance, the importance of the combination between direct and indirect new physics searches has been thoroughly confirmed in the two Higgs doublet models as a representative extended Higgs model~\cite{Kanemura:2014bqa, Kanemura:2014dea, Aiko:2020ksl}. 

The first-order phase transition can be tested by gravitational waves (GW).
In particular, it has been confirmed that the nature of the EWPT can be tested by future GW observations~\cite{Grojean:2006bp, Kakizaki:2015wua, Hashino:2016rvx}. 
The peak frequency of GWs produced by the first-order EWPT is typically around $10^{-3}$\,-\,$10^{-1}\,{\rm Hz}$. 
Such GWs can be observed at future space-based interferometers like LISA~\cite{LISA:2017pwj}, DECIGO~\cite{Kawamura:2011zz}, TianQin~\cite{TianQin:2015yph}, Taiji~\cite{Ruan:2018tsw} and BBO~\cite{Corbin:2005ny}.
The Higgs sector of new physics with the first-order EWPT may be determined precisely by using such GW observations~\cite{Hashino:2018wee}. 

In addition to the GWs, observations of primordial black holes (PBHs) would be also useful to test the first-order EWPT.
Recently, it has been confirmed that PBHs are useful to test first-order phase transitions at the early Universe~\cite{Jung:2021mku,Kawana:2021tde,Hashino:2021qoq,Liu:2021svg,Lu:2022paj,Hashino:2022tcs,Lewicki:2023ioy,Kawana:2022olo,Gouttenoire:2023naa,Jinno:2023vnr,Flores:2024lng,Kanemura:2024pae}. 
For example, the mass of PBHs ($M_{\rm PBH}$) formed by the delayed first-order EWPT is around $10^{-5} M_{\odot}$, where $M_{\odot}$ is the solar mass~\cite{Hashino:2021qoq,Hashino:2022tcs}. 
Such PBHs can be observed in microlensing observations.
The fraction of PBHs ($f_{\rm PBH}$) with $M_{\rm PBH} \sim 10^{-5} M_{\odot}$ has already been constrained up to $f_{\rm PBH} < 10^{-2}$ by Subaru HSC~\cite{Niikura:2017zjd} and OGLE~\cite{Niikura:2019kqi}. 
In future microlensing observations such as PRIME~\cite{Kondo_2023} and Roman telescope~\cite{fardeen2023astrometric}, it is expected that stronger constraints on $f_{\rm PBH}$ can be obtained. It indicates that we can test the first-order EWPT by using current and future PBH observations. 

In this paper, we employ an effective field theoretical framework called the nearly aligned Higgs Effective Field Theory (naHEFT). 
There are two appropriate EFT frameworks: the Standard Model Effective Field Theory (SMEFT)~\cite{Buchmuller:1985jz,Hagiwara:1993ck,Grzadkowski:2010es} and the Higgs Effective Field Theory (HEFT)~\cite{Feruglio:1992wf,Alonso:2012px,Brivio:2013pma,Buchalla:2013rka,Buchalla:2017jlu,Falkowski:2019tft,Cohen:2020xca,Sun:2022ssa,Sun:2022snw}.
The SMEFT can well describe decoupling new physics. 
On the other hand, the HEFT can describe not only decoupling new physics models but also non-decoupling new physics models (see e.g., Refs.~\cite{Banta:2021dek, Banta:2022rwg, Buchalla:2023hqk}). 
The naHEFT is based on the HEFT with the assumption: all deviations in the Higgs boson couplings are originated from quantum effects of new particles which are integrated out~\cite{Kanemura:2022txx,Kanemura:2021fvp}.
This assumption is taken due to the fact that the measured Higgs couplings are currently consistent with the SM prediction as we mentioned above.
If new particles integrated out have the same non-decoupling property, physics of the electroweak symmetry breaking can be then described by several parameters in the naHEFT, so that there is a correlation among the Higgs boson couplings such as $h \gamma \gamma$, $hWW$ and $hhh$ couplings.
In this paper, we analyze the EWPT satisfying the sphaleron decoupling condition in Eq.\,\eqref{eq:vnTn} and the completion condition of the phase transition~\cite{Turner:1992tz}, and investigate the relation among the deviations in the Higgs boson couplings and the dynamics of the EWPTs. 
We also take into account the GW spectrum~\cite{Grojean:2006bp}, as well as the PBHs predicted at the EWPT~\cite{Hashino:2021qoq,Hashino:2022tcs}.
We show that in new physics models with additional charged scalar fields future precision measurements of the $h \gamma \gamma$ coupling can give a useful prediction on the $hhh$ coupling to realize the strongly first-order EWPT.
Consequently, we confirm that the nature of EWPT and new physics models can be explored by the combination of precision measurements of the Higgs boson couplings at future collider experiments, GW observations at future space-based interferometers and searches for the PBHs.

We outline the structure of our paper. 
In Section~\ref{sec:naHEFT}, the effective Lagrangian in the naHEFT is defined. 
We then calculate the scaling factors for each Higgs coupling by using the Lagrangian.
In Section~\ref{sec:hcoup_GW}, the parameter dependence on the scaling factors in the naHEFT is shown. 
The parameter region explored by future GW observations is also shown. 
In addition, the parameter region where observable amounts of PBHs can be formed is also discussed.
Our conclusion is given in Section~\ref{sec:conclusion}.

%%%%%%%%%%%%%%%%%%%%%%%%%%%%%%%%%%%%%%%%

\section{Lagrangian in the nearly aligned Higgs EFT \label{sec:naHEFT}}

The effective Lagrangian in the naHEFT at zero temperature is given by 
\begin{align}
\mathcal{L}_{\rm naHEFT} = \mathcal{L}_{\rm SM} + \mathcal{L}_{\rm BSM}, 
\end{align}
where $\mathcal{L}_{\rm SM}$ is the SM Lagrangian. 
The BSM parts without fermion fields are defined by 
\begin{align} 
\label{eq:LnaHEFT_boson}
\mathcal{L}_{\rm BSM} \ni \xi \left[ -\frac{\kappa_{0}}{4} \left[ \mathcal{M}^{2}(h) \right]^{2} \ln \frac{\mathcal{M}^{2}(h)}{\mu^{2}}  + \frac{v^{2}}{2} \mathcal{F}(h) \operatorname{Tr} \left[ D_{\mu} U^{\dagger} D^{\mu} U\right]+\frac{1}{2} \mathcal{K}(h)\partial_{\mu} h \partial^{\mu} h \right], 
\end{align}
with $\xi = 1/(4\pi)^2$. 
The first term represents the BSM correction to the Higgs potential. 
The parameters $\kappa_{0}$ and $\mu^2$ correspond to the degrees of freedom of new particles and the renormalization scale, respectively.
$h$ is the observed Higgs field, and $v$ is the vacuum expectation value, which is taken as $v  = 246\,{\rm GeV}$.
The matrix $U$, which describes the Nambu-Goldstone bosons $(\pi^{\pm}, \pi^{3})$, is defined as 
\begin{align}
U = \exp \left[ i \frac{\pi^{a} \tau^{a}}{v} \right], \quad \pi^{\pm} = \frac{1}{\sqrt{2}} (\pi^{1} \mp i \pi^{2}), 
\end{align}
where $\tau^{a} \, (a =1, \, 2, \,3)$ are the Pauli matrices. 
The covariant derivative of the matrix $U$ is defined by 
\begin{align}
D_{\mu} U = \partial_{\mu} U + i g \mathbf{W}_{\mu} U -  i g'  U \mathbf{B}_{\mu}, 
\end{align}
where the $SU(2)_{L}$ and $U(1)_{Y}$ gauge fields are given by $\mathbf{W}_{\mu} = \sum_{a=1}^{3} W_{\mu}^{a} \frac{\tau^{a}}{2}$ and $\mathbf{B}_{\mu} = B_{\mu} \frac{\tau^{3}}{2}$, respectively. 
The factor $g$ and $g'$ are the $SU(2)_{L}$ and $U(1)_{Y}$ gauge couplings. 
The three form factors $\mathcal{M}^2(h), \, \mathcal{F}(h)$ and $\mathcal{K}(h)$ parameterize new physics effects. 

It can be confirmed that Eq.~\eqref{eq:LnaHEFT_boson}, which only includes the leading terms in the Higgs EFT, does not account for new physics effects in loop-induced Higgs couplings such as $h \gamma \gamma$ and $h Z \gamma$. 
Therefore, we consider the following effective Lagrangian
\begin{align} 
\begin{aligned}
\label{eq:LnaHEFT_boson}
\mathcal{L}_{\rm BSM} \ni  ~\xi &\left[ -\frac{\kappa_{0}}{4} \left[ \mathcal{M}^{2}(h) \right]^{2} \ln \frac{\mathcal{M}^{2}(h)}{\mu^{2}}  + \frac{v^{2}}{2} \mathcal{F}(h) \operatorname{Tr} \left[ D_{\mu} U^{\dagger} D^{\mu} U\right]+\frac{1}{2} \mathcal{K}(h)\partial_{\mu} h \partial^{\mu} h \right. \\ 
& \left.
~ + g^2 \mathcal{F}_{W}(h) \operatorname{Tr} [{\bf W}_{\mu \nu} {\bf W}^{\mu \nu}] 
+ g'^2\mathcal{F}_{B}(h) \operatorname{Tr} [{\bf B}_{\mu \nu} {\bf B}^{\mu \nu}] 
\right. \\ & \left.
~- g g'\mathcal{F}_{BW}(h) \operatorname{Tr} [U {\bf B}_{\mu \nu} U^{\dagger} {\bf W}^{\mu \nu}]
\right], 
\end{aligned}
\end{align}
where ${\bf W}_{\mu \nu}$ and ${\bf B}_{\mu \nu}$ are the field strength of the $SU(2)_{L}$ and $U(1)_{Y}$ gauge fields, respectively. 
The new three form factors $ \mathcal{F}_{i}(h) \, (i =W, \, B, \, BW)$ parameterize new physics effects in the $h \gamma \gamma$ and $h Z \gamma$ couplings. 
The effective Lagrangian in Eq.\,\eqref{eq:LnaHEFT_boson} is valid if the new physics scale is higher than the electroweak scale. 

In this paper, we assume that the form factor $\mathcal{M}^2(h)$ is characterized by two parameters $r$ and $\Lambda$ as
\begin{align}
\mathcal{M}^2(h) = (1 - r) \Lambda^2 + r \Lambda^2\left(1 + \frac{h}{v} \right)^2, 
\end{align}
where $r$ is the non-decoupingness which parameterizes the decoupling behavior~\cite{Kanemura:2021fvp}. 
We assume that all BSM particles have the same property for simplicity.
It is also assumed that the value of $r$ lies within the region $0 \leq r \leq 1$. 
The parameter $\Lambda$ is the scale of new physics. 
When $r \simeq 1$, the mass of new particles comes from the Higgs vacuum expectation value (VEV). 
This case corresponds to the non-decoupling case. 
On the other hand, if $r \simeq 0$, the mass of new particles does not depend on the Higgs VEV. 
Such new physics exhibits the decoupling property. 

In addition to $\mathcal{M}^2(h)$, we assume that the form factors $\mathcal{K}(h)$ and $\mathcal{F}(h)$ are given by
\begin{align}
\mathcal{K}(h) = \kappa_{0} \frac{\Lambda^2}{3v^2} r \left[ 1 - (1- r) \frac{\Lambda^2}{\mathcal{M}^2(h)} \right], \quad \mathcal{F}(h) = 0. 
\label{eq:formKF}
\end{align}
If electric charges of new particles are less than or equal to 2, $\kappa_{0}$ is expressed by $\kappa_{0} = n_{0} + 2 n_{+} + 2 n_{++}$, where $n_{0}$, $n_{+}$ and $n_{++}$ are the number of new neutral, singly and doubly charged particles, respectively. 
It has been confirmed that the above form factors can well describe extended Higgs models like two Higgs doublet models~\cite{Aiko:2020ksl} and Higgs singlet models~\cite{Kanemura:2015mxa,Kanemura:2015fra}. 
If $\mathcal{K}(h)$ takes a non-zero value, we must define a canonical Higgs field $(\hat{h})$ to normalize the kinetic term of the Higgs field~\cite{Kanemura:2021fvp}. 
In the following, we express $\hat{h}$ as $h$ for simplicity. 

It is well known that the sub-leading terms can describe the non-decoupling effects in $h \gamma \gamma$ and $h Z \gamma$ couplings~\cite{Herrero:2020dtv,Arco:2023sac,Buchalla:2023hqk}. 
The expression of the form factors $ \mathcal{F}_{i}(h)$ has been discussed in the non-decoupling limit~\cite{Buchalla:2023hqk}. 
If the mass of particles is sufficiently heavier than the Higgs mass, the form factors should reproduce the results of the low-energy theorem~\cite{Shifman:1979eb}. 
We assume that the expressions of $ \mathcal{F}_{i}(h)$ are given by 
\begin{align}
\mathcal{F}_{W}(h) = \mathcal{F}_{B}(h) = \frac{b}{2} \ln \left[ \tilde{r} \left( 1 + \frac{h}{v} \right)^2 + 1- \tilde{r} \right], ~ \mathcal{F}_{BW}(h) = 0, 
\label{eq:FWB}
\end{align}
where the parameters $b$ and $\tilde{r}$ characterize deviations in the $h \gamma \gamma$ and $h Z \gamma$ couplings. 
The parameter $b$ depends on the type and degrees of freedom of new particles. 
If we consider an extension of the SM with new singly and doubly charged particles, the parameter $b$ is given by $b = (n_{+} + 4 n_{++})/3$. 
The parameter $\tilde{r}$ characterizes the non-decouplingness of charged particles. 
We here assumed that all new charged particles have the same non-decouplingness. 
Although $r$ and $\tilde{r}$ are not the same in general, we assume $\tilde{r} = r$ in the following for simplicity. 
When the decoupling limit ($r \sim 0$), the form factors $\mathcal{F}_{i}(h)$ vanish. 
On the other hand, in the non-decoupling situation where $r \sim 1$, Eq.~\eqref{eq:FWB} reproduces the results that are consistent with the low-energy theorem~\cite{Shifman:1979eb}. 

We define the scaling factors describing the deviations in Higgs couplings from the SM prediction as 
\begin{align}
\kappa_{V} = \frac{g_{hVV}^{\rm EFT}}{g_{hVV}^{\rm SM}}, \quad 
\kappa_{f} = \frac{y_{hff}^{\rm EFT}}{y_{hff}^{\rm SM}}, \quad
\kappa_{3} = \frac{\lambda_{hhh}^{\rm EFT}}{\lambda_{hhh}^{\rm SM}}, 
\end{align}
where $g_{hVV}^{\rm SM}\,(V = W, \, Z)$, $y_{hff}^{\rm SM}$ and $\lambda_{hhh}^{\rm SM}$ are the $hVV$, $hf \bar{f}$ and $hhh$ couplings in the SM, respectively. 
The couplings $g_{hVV}^{\rm EFT}$, $y_{hff}^{\rm EFT}$ and $\lambda_{hhh}^{\rm EFT}$ are those in the naHEFT.  
For the scaling factors of loop-induced Higgs couplings are defined by
\begin{align}
\kappa_{\gamma \gamma}^2 = \frac{ \Gamma_{h \to \gamma \gamma}^{\rm EFT} }{\Gamma_{h \to \gamma \gamma}^{\rm SM} }, \quad
\kappa_{Z \gamma}^2 = \frac{ \Gamma_{h \to Z \gamma}^{\rm EFT} }{\Gamma_{h \to Z \gamma}^{\rm SM} }, 
\end{align}
where $\Gamma_{h \to XY}^{\rm SM}$ and $\Gamma_{h \to XY}^{\rm EFT}$ are the decay rates of the mode $h \to XY$ in the SM and in the naHEFT, respectively. 

The scaling factors can be expressed in terms of the parameters in the naHEFT as 
\begin{align}
\begin{aligned}
&\kappa_{V} = \kappa_{f} =  1 - \kappa_{0} \frac{\xi}{6} \frac{\Lambda^2}{v^2} r^2, \\
&\kappa_{3} = 1 + \kappa_{0} \frac{4 \xi}{3} \frac{\Lambda^4}{v^2 m_{h}^2} \left[ r^3 - \frac{m_{h}^2}{8 \Lambda^2} r^2 (3-2r)\right], \\
&\kappa_{\gamma \gamma}^2 \simeq \left| \kappa_{V} -  \frac{br}{F_{\rm SM}}  \right|^2, \\
&\kappa_{Z \gamma}^2 \simeq \left| \kappa_{V} -  \frac{br}{G_{\rm SM}} \left( J_{3}^{\rm new} - s_{W}^2 \right)  \right|^2, 
\label{eq:scaling_factors}
\end{aligned}
\end{align} 
where $F_{\rm SM} = 6.492$, $G_{\rm SM} = 11.65$, and $s_{W}$ is the Weinberg angle. 
The parameter $J_{3}^{\rm new}$ parameterizes the ratio between the third isospin component and the electrical charge of new particles.
The expressions for $\kappa_{V}$ and $\kappa_{3}$ are the same as the results in Ref.~\cite{Kanemura:2021fvp}. 
In the following, we utilize the above scaling factors in discussing phenomenological results. 
We note that constraints from $h Z \gamma$ and $hf\bar{f}$ measurements are not considered in this paper because those are weaker than the current constraints from the $hVV$ and the $h \gamma \gamma$ coupling measurements~\cite{CMS:2022dwd, ATLAS:2022vkf}.

We here define the effective potential at finite temperatures in the naHEFT. 
The effective potential including thermal effects in the naHEFT is given by~\cite{Kanemura:2022txx}
\begin{align}
V_{\rm naHEFT}(\phi, T) = V_{\rm SM}(\phi, T) + V_{\rm BSM}(\phi, T), 
\label{eq:VEFT}
\end{align}
where $V_{\rm SM}(\phi, T)$ and $V_{\rm BSM}(\phi, T)$ are the SM part and the BSM part. 
For the SM part, we include the contributions from the electroweak gauge fields, the SM Higgs field and top quark. 
We use the formula shown in Ref.~\cite{Kanemura:2022txx}.

Predictions on GW spectra produced by the first-order EWPT can be discussed by using the effective potential in Eq.~\eqref{eq:VEFT}. 
The GWs produced by the first-order EWPT have three sources: bubble collisions, compression waves (sound waves) and magnetohydrodynamics turbulence. 
Then, the GW spectra $\Omega_{\rm GW}(f)$ can be expressed by 
\begin{align}
h^2 \Omega_{\rm GW}(f) \simeq h^2 \Omega_{\varphi}(f) + h^2 \Omega_{\rm sw}(f) +  h^2 \Omega_{\rm turb}(f), 
\end{align}
where $\Omega_{\varphi}, \, \Omega_{\rm sw}$ and $\Omega_{\rm turb}$ are the bubble collision part, the sound wave part and the turbulence part, respectively. 
We utilize the fitting functions for the GW spectrum $ \Omega_{\rm GW}(f)$ shown in Ref.~\cite{Caprini:2015zlo}. 
To discuss the detectability of the GWs, we evaluate the signal-to-noise ratio~\cite{Caprini:2015zlo}
\begin{align}
{\rm SNR} = \sqrt{ \mathcal{T} \int^{f_{\rm max}}_{f_{\rm min}} df \left[ \frac{h^2 \Omega_{\rm GW}(f)}{h^2 \Omega_{\rm sens}(f)} \right]^2 }, 
\label{eq:SNR}
\end{align}
where $h^2 \Omega_{\rm sens}(f)$ and $\mathcal{T}$ are the sensitivity and the observation duration of future space-based interferometers like LISA~\cite{LISA:2017pwj} and DECIGO~\cite{Kawamura:2011zz}. 
We take $\mathcal{T} = 1.26 \times 10^{8}\,{\rm s}$ for all GW experiments. 
In this paper, we conclude that the GWs can be detected if the condition ${\rm SNR}>10$ is satisfied~\cite{Cline:2021iff}.

We also take into account PBH observations.
We focus on the scenario where the delay of the first-order EWPT realizes the large density fluctuation to form the PBHs~\cite{Liu:2021svg}. 
In the following, we perform the same analyses as in Ref.~\cite{Hashino:2022tcs}.

\section{Higgs couplings and first-order EWPT in the naHEFT \label{sec:hcoup_GW}}

We here discuss predictions on Higgs couplings by using the results in the previous section. 
In addition, we analyze the properties of the GWs and PBHs produced by the first-order EWPT. 

\begin{figure}[t]
\centering
\includegraphics[width=0.98\textwidth]{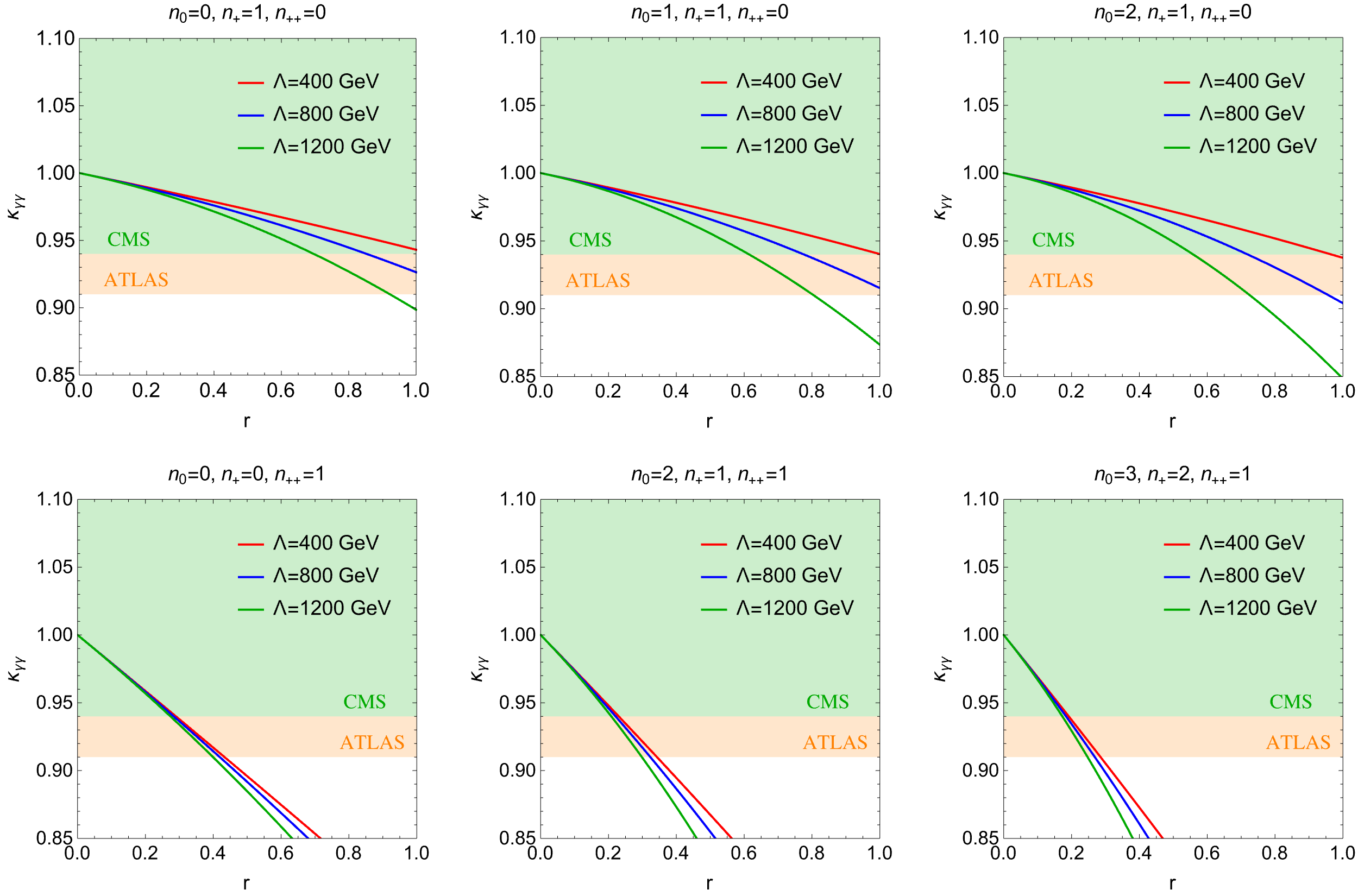}
\caption{
The relation between $\kappa_{\gamma \gamma}$ and the non-decouplingness $r$. 
The parameter $\Lambda$ corresponds to the mass scale of new particles.
The green and orange regions indicate the current allowed regions from the CMS result~\cite{CMS:2022dwd} and the ATLAS result~\cite{ATLAS:2022tnm}, respectively.
The parameters $n_{0}$, $n_{+}$ and $n_{++}$ are the degrees of freedom of neutral, singly charged and doubly charged particles, respectively.
\label{fig:kg_r}
}
\end{figure}

In Fig.\,\ref{fig:kg_r}, the relation between $\kappa_{\gamma \gamma}$ and the non-decouplingness $r$ is shown. 
The current constraints from the ATLAS and CMS are shown in Fig.\,\ref{fig:kg_r} as the orange and green regions, respectively. 
The constraint on $\kappa_{\gamma \gamma}$ from the CMS results is $\kappa_{\gamma \gamma} = 1.10 \pm 0.16$ at $95\,\%$ confidence level~\cite{CMS:2022dwd}. 
For the ATLAS, the current constraint is $\kappa_{\gamma \gamma} = 1.02 \pm 0.11$ at $95\,\%$ confidence level, where it is assumed that the $hgg$ coupling is the same as the SM prediction~\cite{ATLAS:2022tnm}. 
Fig.\,\ref{fig:kg_r} indicates that the large $\Lambda$ can make the deviation in $\kappa_{\gamma \gamma}$ large. 
The top three figures indicate that the deviation in $\kappa_{\gamma \gamma}$ can be large if we take into account charge neutral particles. 
This effect comes from the deviation in $\kappa_{V}$.
The bottom three figures imply that extended Higgs models with doubly charged scalar bosons are strongly constrained by $h \gamma \gamma$ coupling measurements. 
These results are consistent with the previous results in the specific renormalizable models~\cite{Arhrib:2011vc, Aoki:2012yt, Chiang:2017vvo,Chiang:2018xpl}.
The result in Fig.\,\ref{fig:kg_r} indicates that the value of $\kappa_{\gamma \gamma}$ becomes SM-like when one takes the decoupling limit $r \sim 0$. 
This behavior is consistent with the decoupling theorem~\cite{Appelquist:1974tg}. 
On the other hand, the relatively large non-decouplingness can be constrained by the $h \gamma \gamma$ coupling measurements at current and future collider experiments. 

\begin{figure}[t]
\centering
\includegraphics[width=0.98\textwidth]{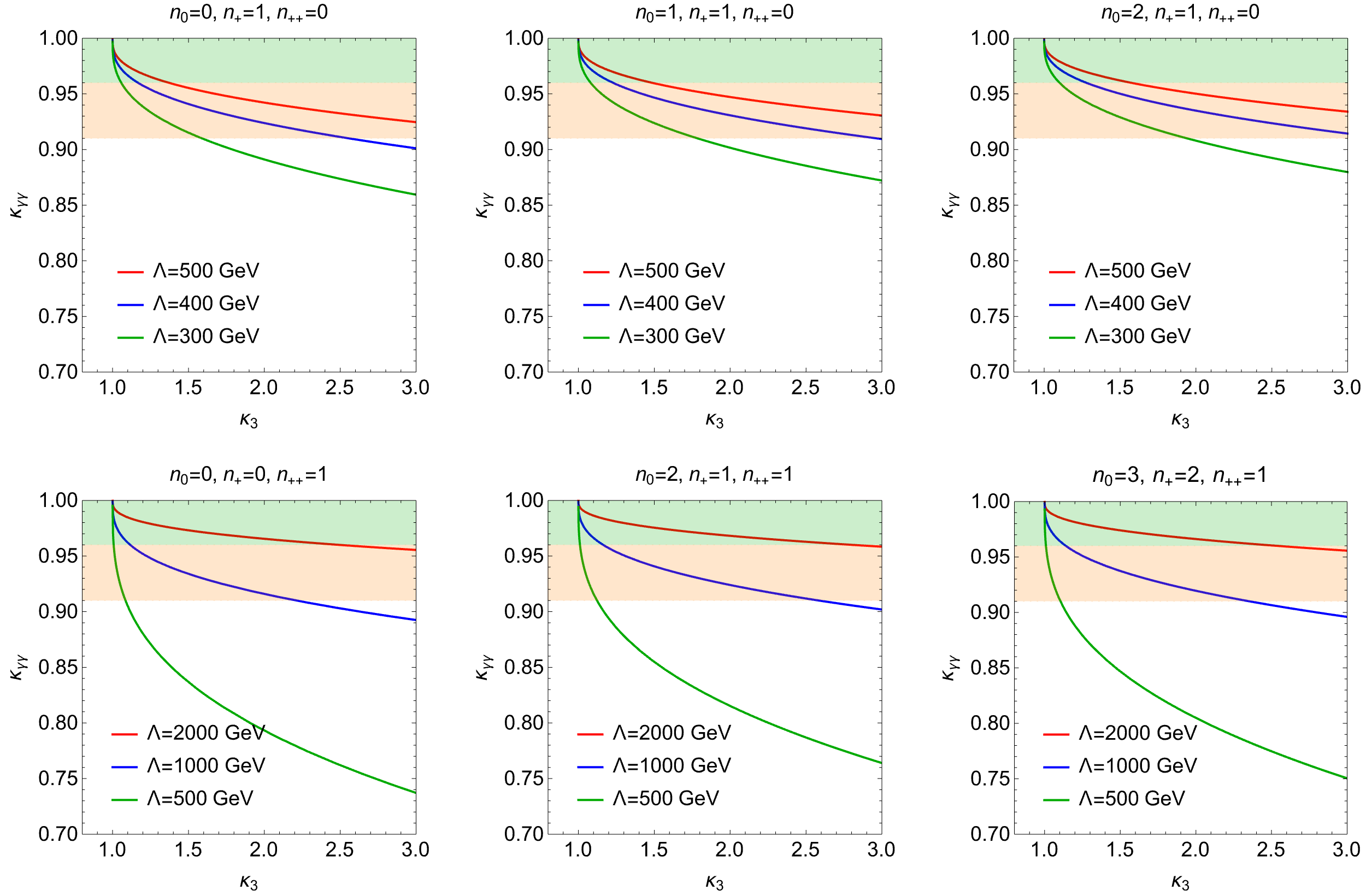}
\caption{
Correlation between $\kappa_{3}$ and $\kappa_{\gamma \gamma}$ for each benchmark point. 
The parameter $\Lambda$ corresponds to the mass scale of new particles.
The green and orange regions indicate the current allowed regions from the CMS result~\cite{CMS:2022dwd} and the ATLAS result~\cite{ATLAS:2022tnm}, respectively.
\label{fig:kg_k3}
}
\end{figure}

In Fig.\,\ref{fig:kg_k3}, the correlation between $\kappa_{3}$ and $\kappa_{\gamma \gamma}$ is shown. 
The figure indicates that large $hhh$ and small $h \gamma \gamma$ couplings can be realized if the scale of new physics is relatively high. 
We note that the $h \gamma \gamma$ coupling may be measured with about $1\,\%$ accuracy by combining results at the HL-LHC with that at the ILC with $\sqrt{s} = 250\,{\rm GeV} $ at $68\,\%$ confidence level~\cite{Bambade:2019fyw}. 
As we confirm later, the condition in Eq.~\eqref{eq:vnTn} requires a large deviation in the $hhh$ coupling. 
Therefore, the new physics scale should be large to satisfy the condition in Eq.~\eqref{eq:vnTn} and future precise $h \gamma \gamma$ measurement results simultaneously.

\begin{figure}[t]
\centering
\includegraphics[width=0.325\textwidth]{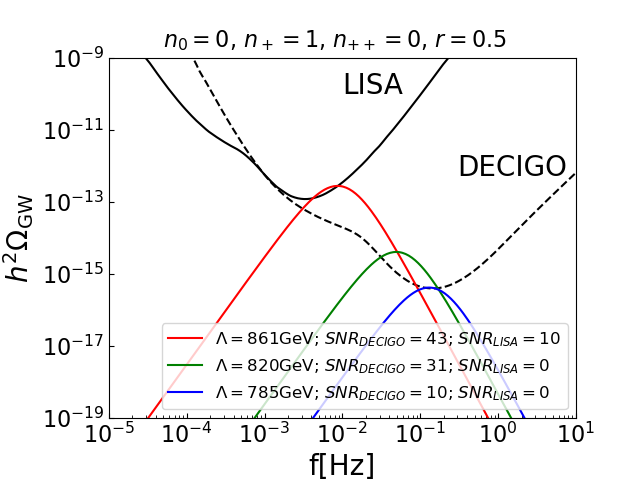}
\includegraphics[width=0.325\textwidth]{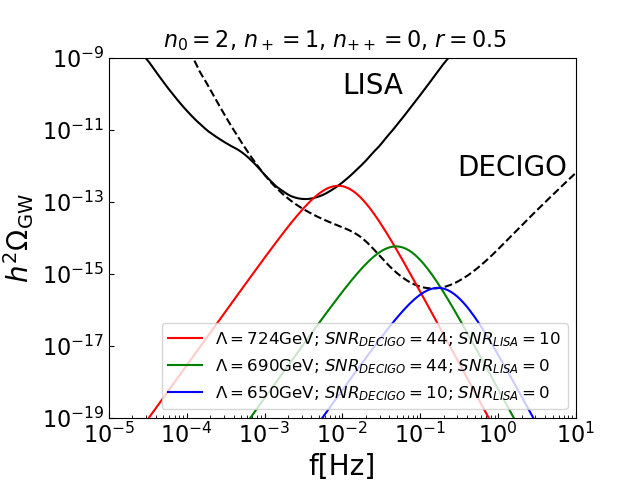}
\includegraphics[width=0.325\textwidth]{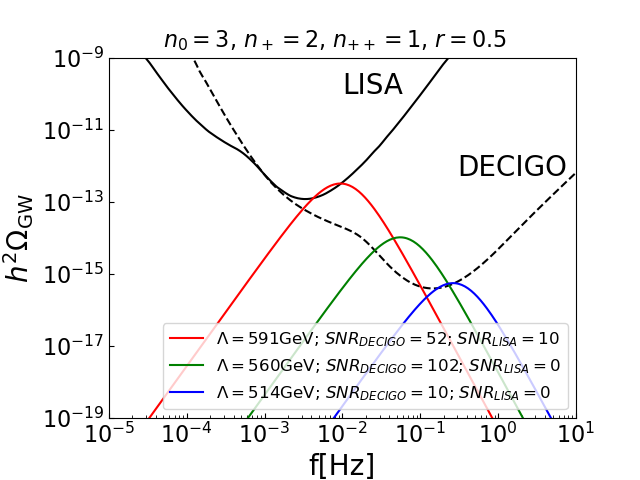}
\caption{
GW spectra in the naHEFT for each benchmark point. 
The parameters $r$ and $\Lambda$ correspond to the non-decouplingness and the mass scale of new particles, respectively.
The black solid and dashed lines are the peak integrated sensitivity curves for LISA and DECIGO, respectively~\cite{Cline:2021iff}. 
If the peak of GW spectra is higher than the sensitivity curves, the detectability criterion ${\rm SNR}>10$ is satisfied. \label{fig:GW}}
\end{figure}

In Fig.\,\ref{fig:GW}, the prediction on GW spectra in three benchmark points is shown. 
The black solid and dashed lines are the peak integrated sensitivity curves for LISA and DECIGO, respectively~\cite{Cline:2021iff}. 
If the peak position of GW spectra is higher than the sensitivity curves, the detectability criterion ${\rm SNR}>10$ is satisfied. 
Therefore, the blue benchmark point can be tested by LISA. 
On the other hand, the orange and green benchmark points can be explored by DECIGO. 
We here assume that the wall velocity is $95\,\%$ of the speed of light. 

\begin{figure}[t]
\centering
\includegraphics[width=0.32\textwidth]{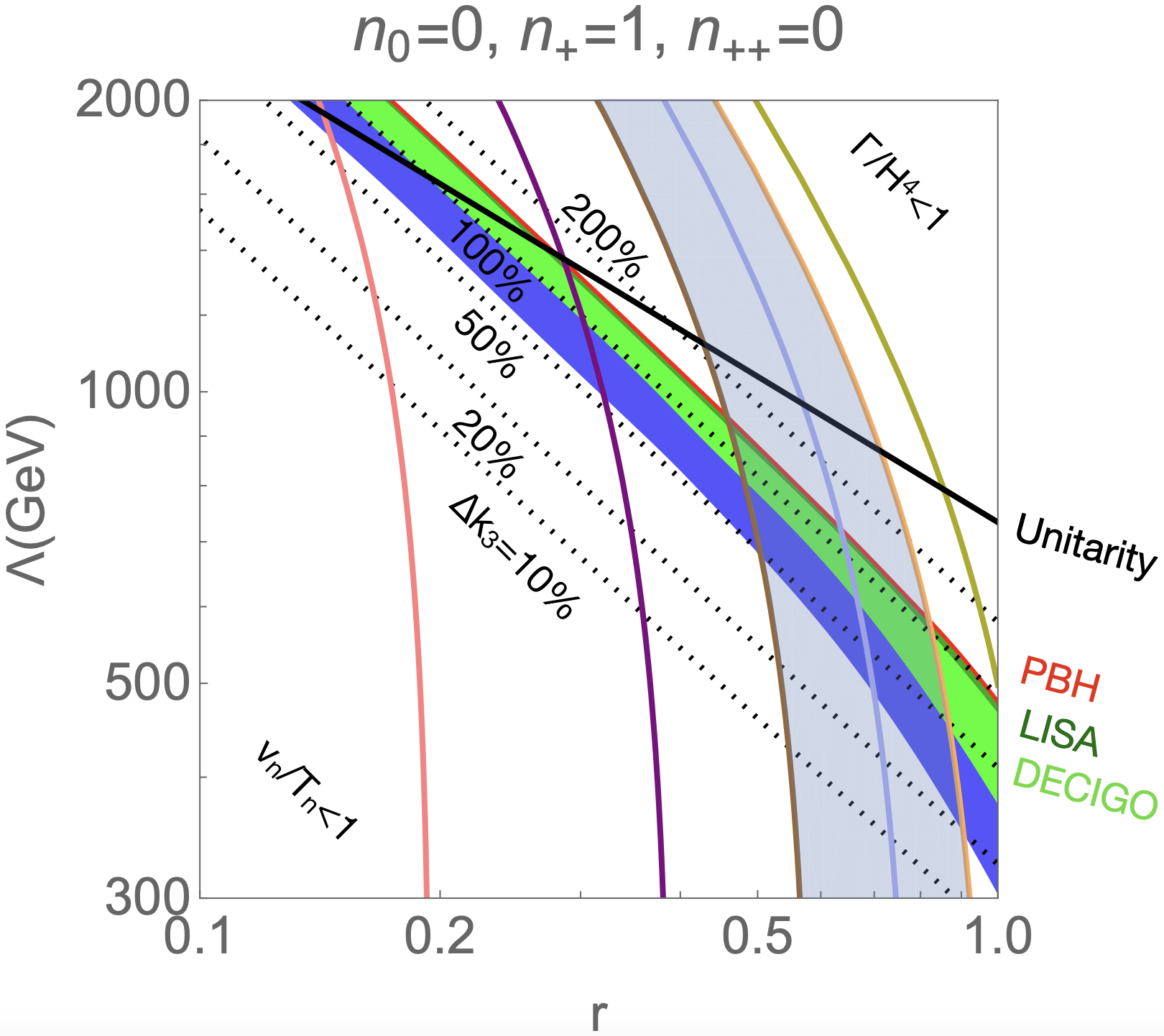}
\includegraphics[width=0.32\textwidth]{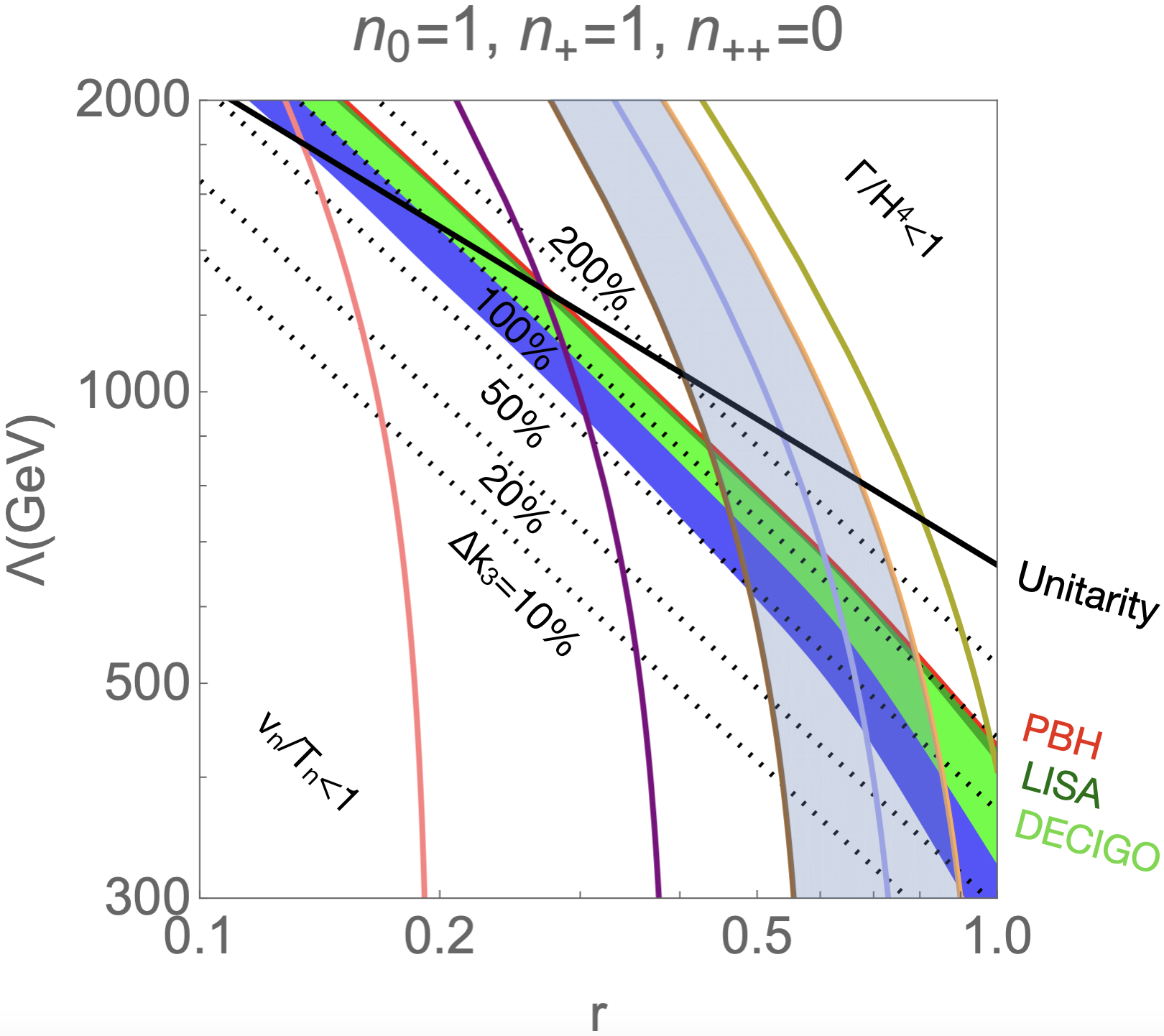}
\includegraphics[width=0.32\textwidth]{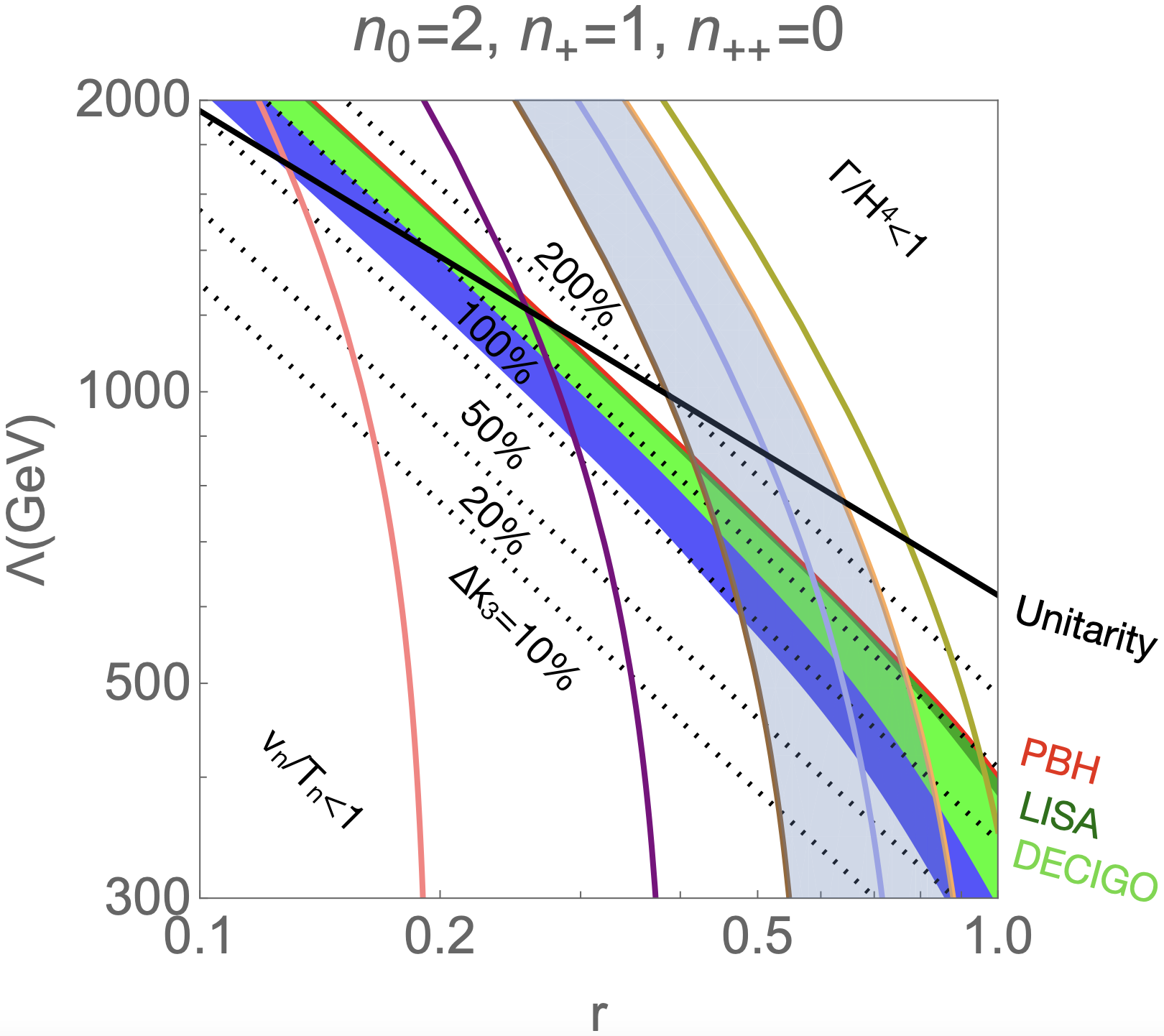}\\
\includegraphics[width=0.32\textwidth]{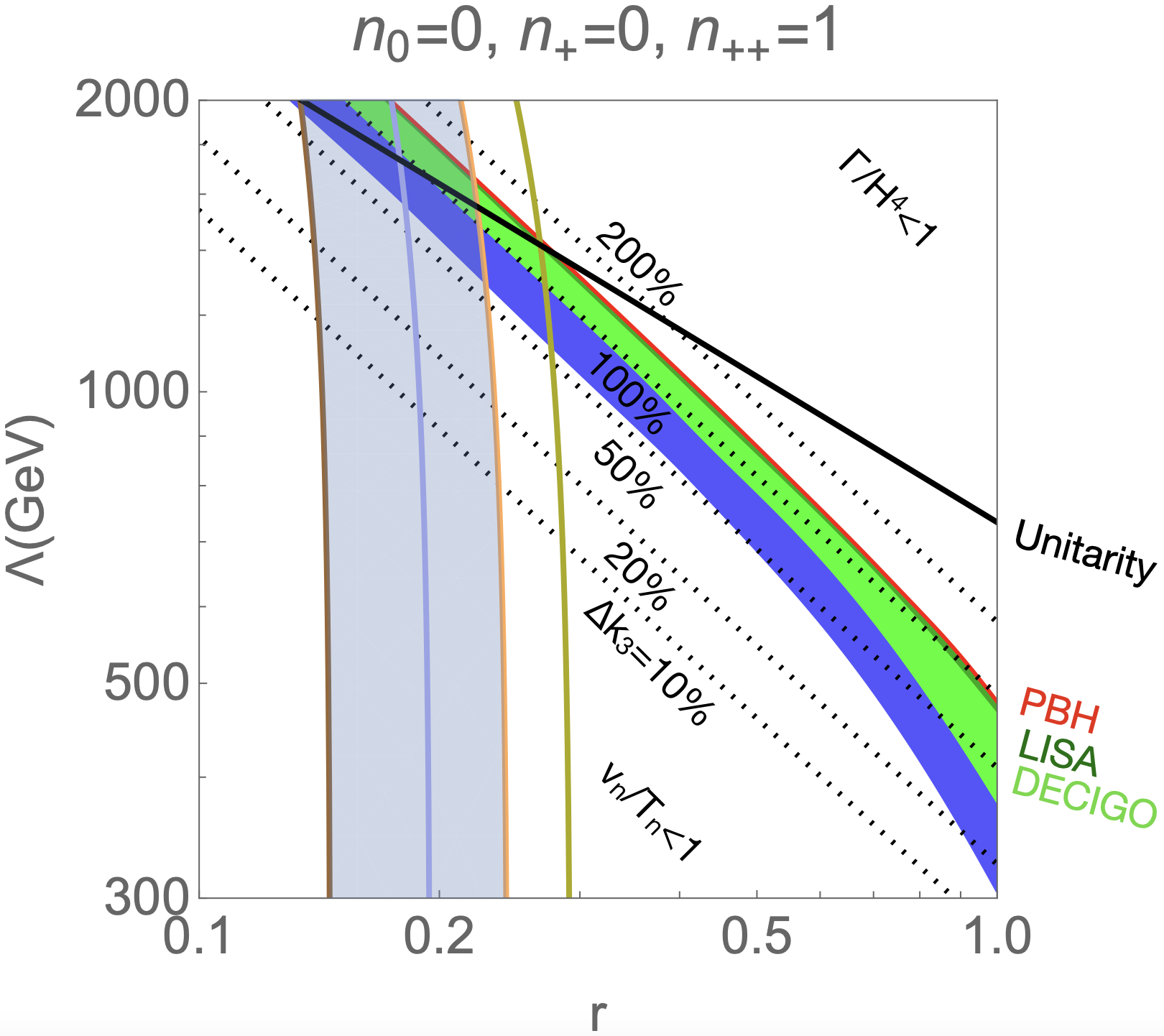}
\includegraphics[width=0.32\textwidth]{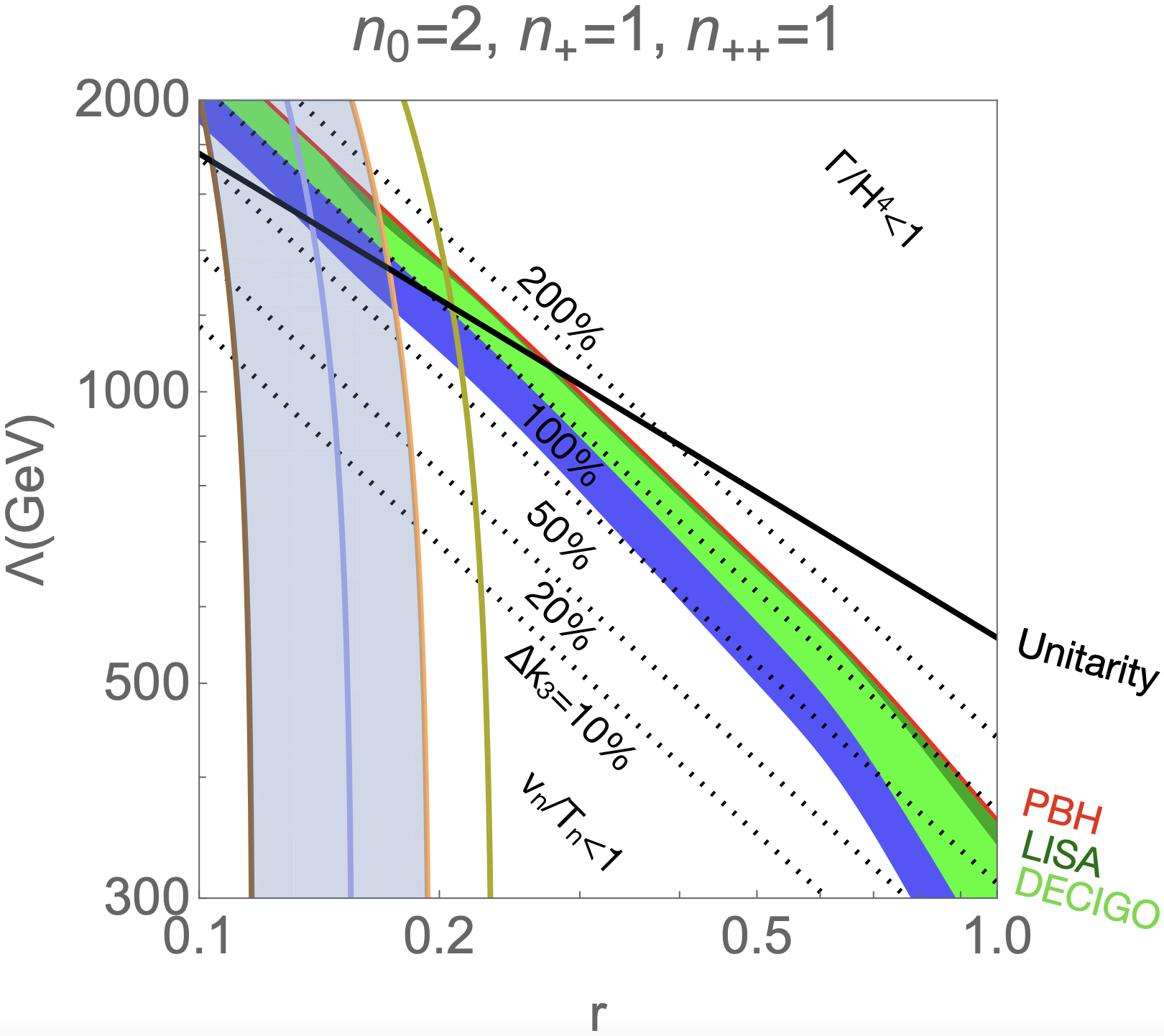}
\includegraphics[width=0.32\textwidth]{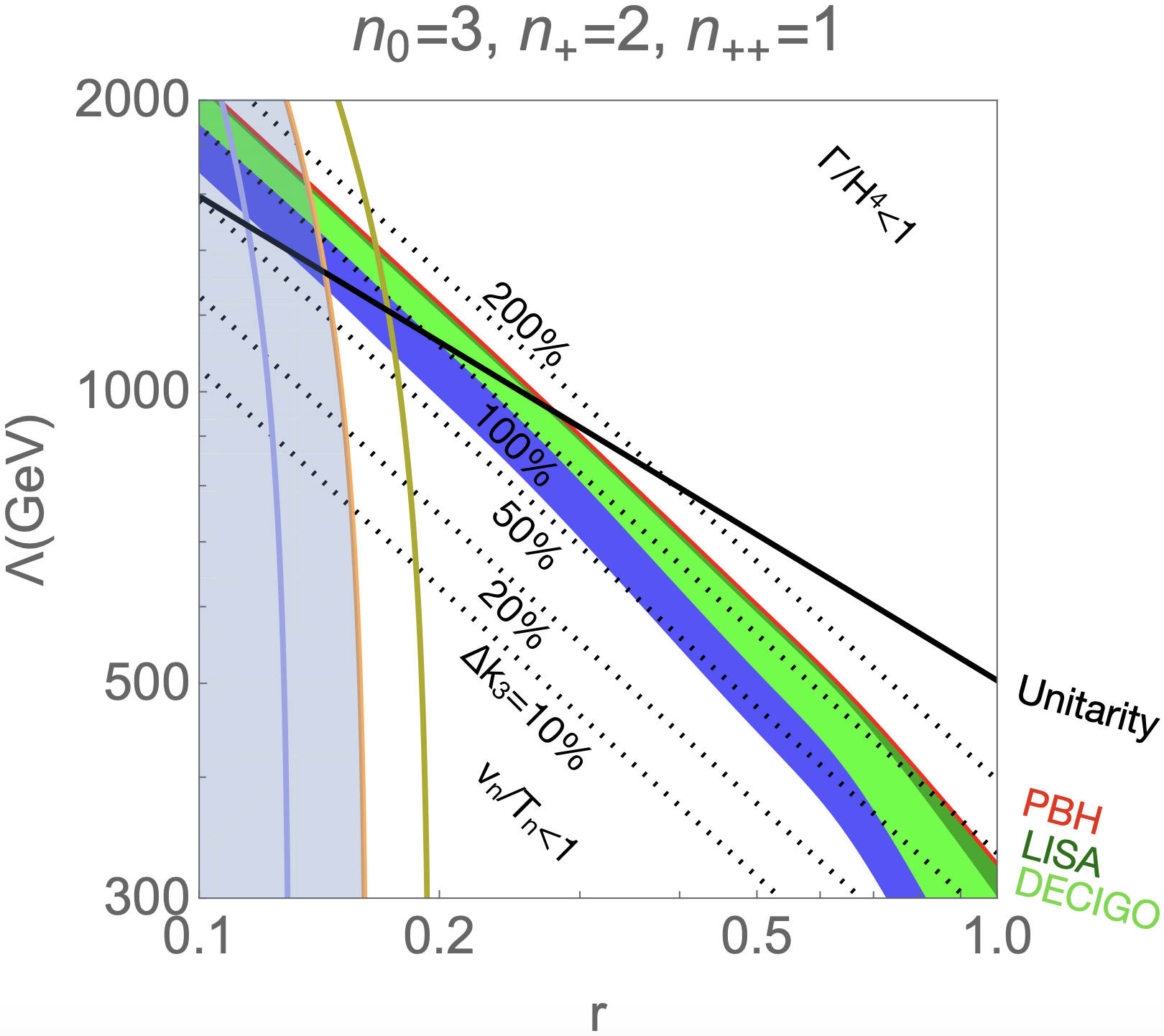}\\
\vspace{0.2cm}
\includegraphics[width=0.75\textwidth]{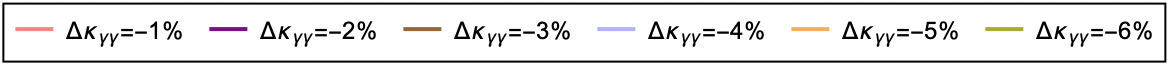}
\caption{
Parameter region explored by $h \gamma \gamma$, $hhh$ and GW observations. 
The parameters $r$ and $\Lambda$ correspond to the non-decouplingness and the mass scale of new particles, respectively.
The values of $\Delta \kappa_{3}$ and $\Delta \kappa_{\gamma \gamma}$ are defined by $\Delta \kappa_{3} \equiv \kappa_{3} - 1$ and  $\Delta \kappa_{\gamma \gamma} \equiv \kappa_{\gamma \gamma} - 1$, respectively. 
The upper right white region is constrained by the perturbative unitarity bound or the completion condition of the phase transition. 
In the lower-left white region, the condition~\eqref{eq:vnTn} is not satisfied. 
The red region can be explored by Subaru HSC, OGLE, PRIME and Roman telescope.
The green and dark green regions are explored by DECIGO and LISA, respectively. 
In the blue region, the first-order EWPT can be basically tested only by collider experiments.
\label{fig:kg_hhh}
}
\end{figure}

In Fig.~\ref{fig:kg_hhh}, the parameter region, which can be explored by $h \gamma \gamma$, $hhh$ and GW observations, is shown for each model.
The colored solid lines are the contours of the $h \gamma \gamma$ coupling. 
The black dotted lines are contours of the deviation in the $hhh$ coupling. 
In the upper-right white region, the phase transition cannot be finished by today. 
In this paper, we define the completion condition of the phase transition by using $\Gamma/H^4$, where $\Gamma$ and $H$ are the nucleation rate and Hubble parameter. 
If $\Gamma/H^4$ cannot be larger than unity for any temperatures, we conclude that such phase transitions cannot be completed by present~\cite{Turner:1992tz}. 
In the lower-left white region, the sphaleron decoupling condition $v_{n}/T_{n} > 1$ is not satisfied. 
The black solid line indicates constraints from the perturbative unitarity bound. 
We here take into account the unitarity bound in the SM with $N$ scalar singlet fields as an example~\cite{Hashino:2016rvx}.
The red region can be explored by current and future microlensing observations such as Subaru HSC~\cite{Niikura:2017zjd}, OGLE~\cite{Niikura:2019kqi}, PRIME~\cite{Kondo_2023} and Roman telescope~\cite{fardeen2023astrometric}. 
This PBH region is the same as the literature~\cite{Hashino:2022tcs}.
The green and dark green regions indicate the parameter region that can be tested at DECIGO and LISA, respectively. 
In the blue region, the first-order EWPT can be basically tested only by collider experiments. 
Fig.~\ref{fig:kg_hhh} indicates that a criterion of the strongly first-order EWPT can be obtained in terms of the $hhh$ coupling by using precise $h \gamma \gamma$ coupling measurements and the condition~\eqref{eq:vnTn}.
Also, Fig.~\ref{fig:kg_hhh} indicates that the large deviation in the $h \gamma \gamma$ coupling is required to realize the strongly first-order EWPT in models with doubly charged particles.

In Table~\ref{table:hhh}, it is summarized that the prediction on the $hhh$ coupling to realize the strongly first-order EWPT when $\Delta \kappa_{\gamma \gamma} = -4\,\% \pm 1\,\%$ for each model.
In the second column labeled ``Required by $\frac{v_{n}}{T_{n}} > 1$ \& $\frac{\Gamma}{H^4} >1$", it is shown that the predictions on the $hhh$ coupling where the EWPT \textit{can be} strongly first-order.
The lower and upper bounds on $\Delta \kappa_{3}$ are determined by $v_{n}/T_{n}>1$ and the completion condition of the phase transition, respectively. 
For the third column labeled ``Conservative bound", it is shown that the predictions on the $hhh$ coupling where the EWPT \textit{is} strongly first-order.
It means that in the bounds shown in the third column, the condition $v_{n}/T_{n}>1$ and the completion condition are safely satisfied simultaneously. 
Thus, if the measured $hhh$ coupling matches the prediction shown in the third column labeled ``Conservative bound", it is expected that the strongly first-order EWPT occurred at the early Universe.  
Therefore, we emphasize that the results in the third column are important in testing the strongly first-order EWPT.

\begin{table}[t]
\begin{tabular}{|c|c|c|c|}
\hline 
~$(n_{0},\, n_{+}, \, n_{++})$~  &  ~Required by $\frac{v_{n}}{T_{n}} > 1$ \& $\frac{\Gamma}{H^4} >1$~ & ~Conservative bound~ & Example of SM extension
\\ \hline 
$(0, \, 1, \, 0)$ & $137\,\%>\Delta \kappa_{3}> 21 \%$ & ~$114\,\%>\Delta \kappa_{3}> 50\,\%$~ & ~A singly charged scalar~
\\ \hline
$(1, \, 1, \, 0)$ & ~$143\,\%>\Delta \kappa_{3}> 19\,\%$~ & $115\,\%>\Delta \kappa_{3}> 47\,\%$ & ~A real triplet scalar~
\\ \hline
$(2, \, 1, \, 0)$ & ~$135\,\%>\Delta \kappa_{3}> 18\,\%$~ & $114\,\%>\Delta \kappa_{3}> 44\,\%$ &~A doublet scalar~
\\ \hline 
$(0, \, 0, \, 1)$ & ~$153\,\%>\Delta \kappa_{3}> 62\,\%$ & $148\,\%>\Delta \kappa_{3}> 65\,\%$ & ~A doubly charged scalar~
\\ \hline
$(2, \, 1, \, 1)$ & ~$160\,\%>\Delta \kappa_{3}> 65\,\%$~ & $150\,\%>\Delta \kappa_{3}> 75\,\%$ & ~A complex triplet scalar~
\\ \hline
$(3, \, 2, \, 1)$ & ~$136\,\%>\Delta \kappa_{3}> 59\,\%$~ & $153\,\%>\Delta \kappa_{3}> 63\,\%$ &~Gerogi-Machacek model~
\\ \hline
\end{tabular}
\caption{
Predictions on the $hhh$ coupling for realizing the strongly first-order EWPT if $\Delta \kappa_{\gamma \gamma} = -4\,\% \pm 1\,\%$ for each model. 
The lower and upper bounds on $\Delta \kappa_{3}$ shown in the second column are determined by $v_{n}/T_{n}>1$ and the completion condition of the phase transition, respectively. 
In the bounds shown in the third column, the condition $v_{n}/T_{n}>1$ and the completion condition are safely satisfied simultaneously. 
If the measured $hhh$ coupling matches the prediction shown in the third column, it is expected that the EWPT is strongly first-order.  
In the final column, corresponding candidates of the SM extensions are shown for each benchmark point.
\label{table:hhh}}
\end{table}

In Ref.~\cite{Kanemura:2022ozv}, it has been confirmed that conservative predictions for the feasibility of the strongly first-order EWPT are given by $\Delta \kappa_{3} > 80\,\%$ in several representative extended Higgs models. 
Our new results indicate that more precise criteria for the strongly first-order phase transition can be obtained by using precise $h \gamma \gamma$ coupling measurements. 
If the $hhh$ coupling measured at future colliders satisfies the prediction shown in Table~\ref{table:hhh}, we can conclude whether the EWPT is strongly first-order or not. 
In addition, we can determine an upper bound on $\Lambda$ by combining precise $h \gamma \gamma$ coupling measurements and the completion condition for the phase transition.
In the case with $(n_{0},\, n_{+}, \, n_{++}) =(0, \, 1, \, 0)$, we can conclude that a new charged particle satisfying $\Lambda < 946\,{\rm GeV}$ and $r>0.457$ should exist if we confirm $\Delta \kappa_{\gamma \gamma} < -3\,\%$.
If we cannot observe any new physics signal at future colliders, we can test the EWPT via the GW and PBH observations in addition to precise Higgs coupling measurements. 

Moreover, Fig.~\ref{fig:kg_hhh} indicates that lower bounds on the $h \gamma \gamma$ coupling can be obtained by combining the condition~\eqref{eq:vnTn} with the perturbative unitarity bound. 
In Table~\ref{table:hgg}, predictions on the $h \gamma \gamma$ coupling are summarized. 
If the measured $h \gamma \gamma$ coupling is larger than these predictions, we expect that the EWPT can be strongly first-order.

\begin{figure}[t]
\centering
\includegraphics[width=0.32\textwidth]{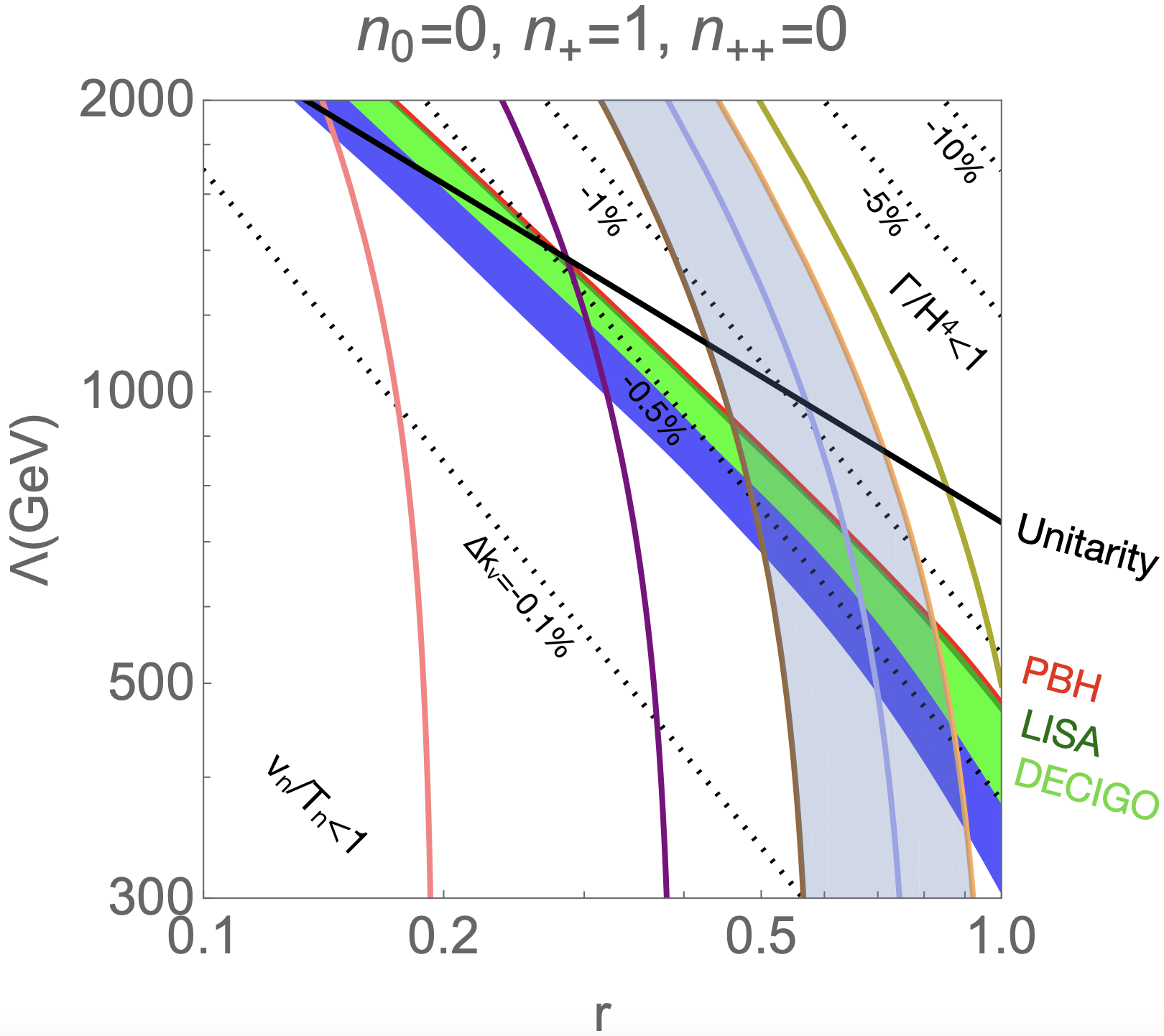}
\includegraphics[width=0.32\textwidth]{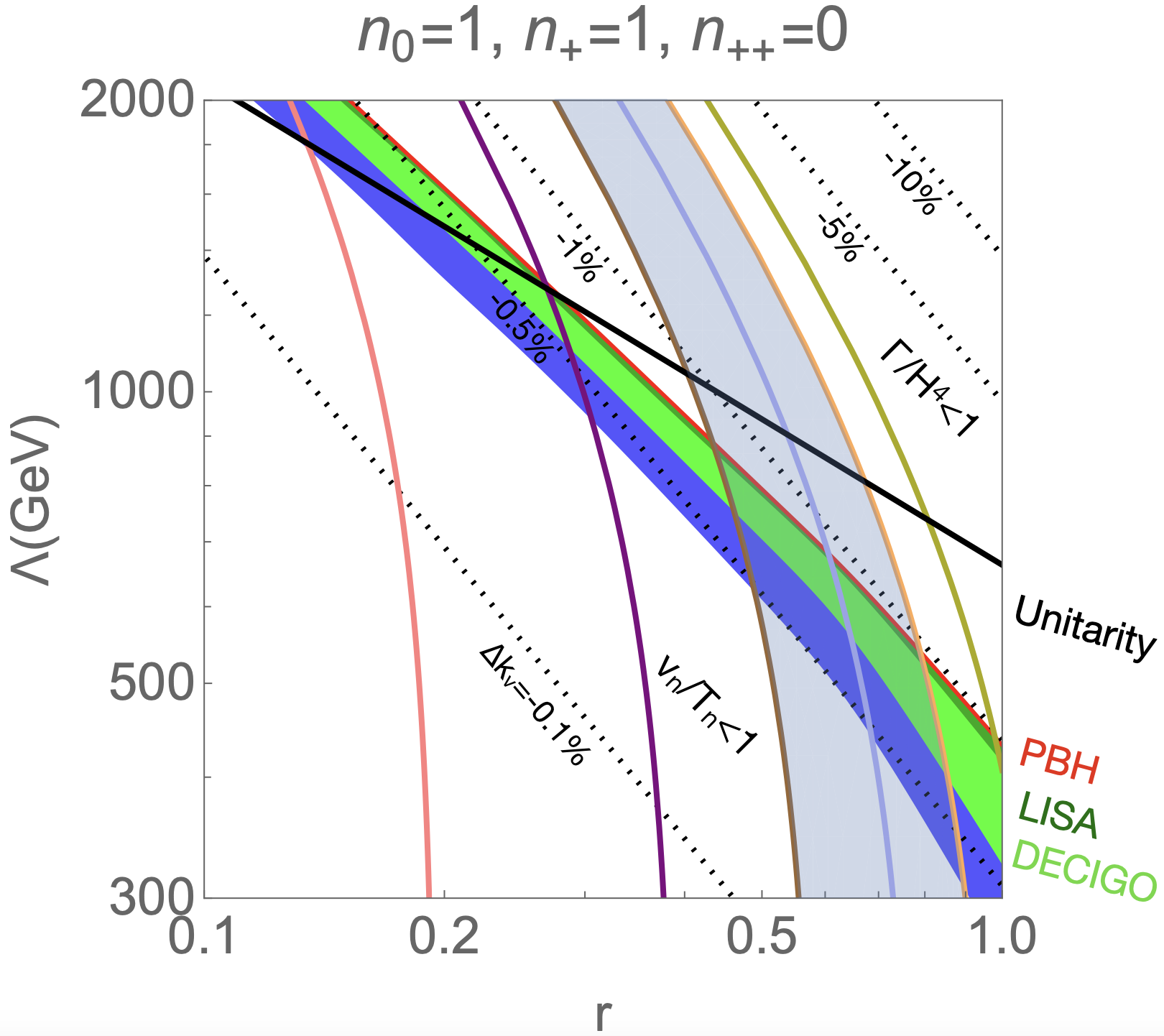}
\includegraphics[width=0.32\textwidth]{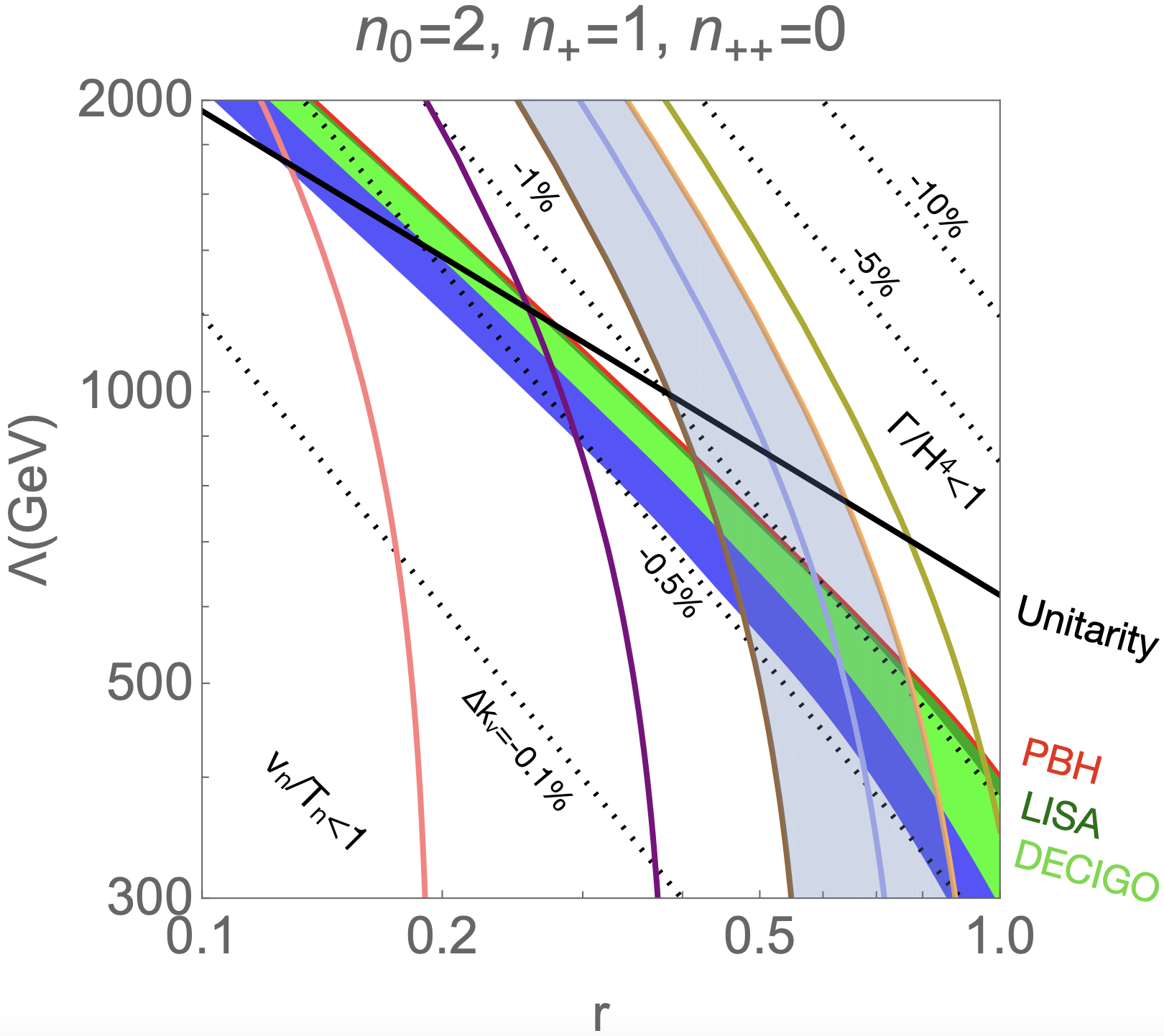}\\
\includegraphics[width=0.32\textwidth]{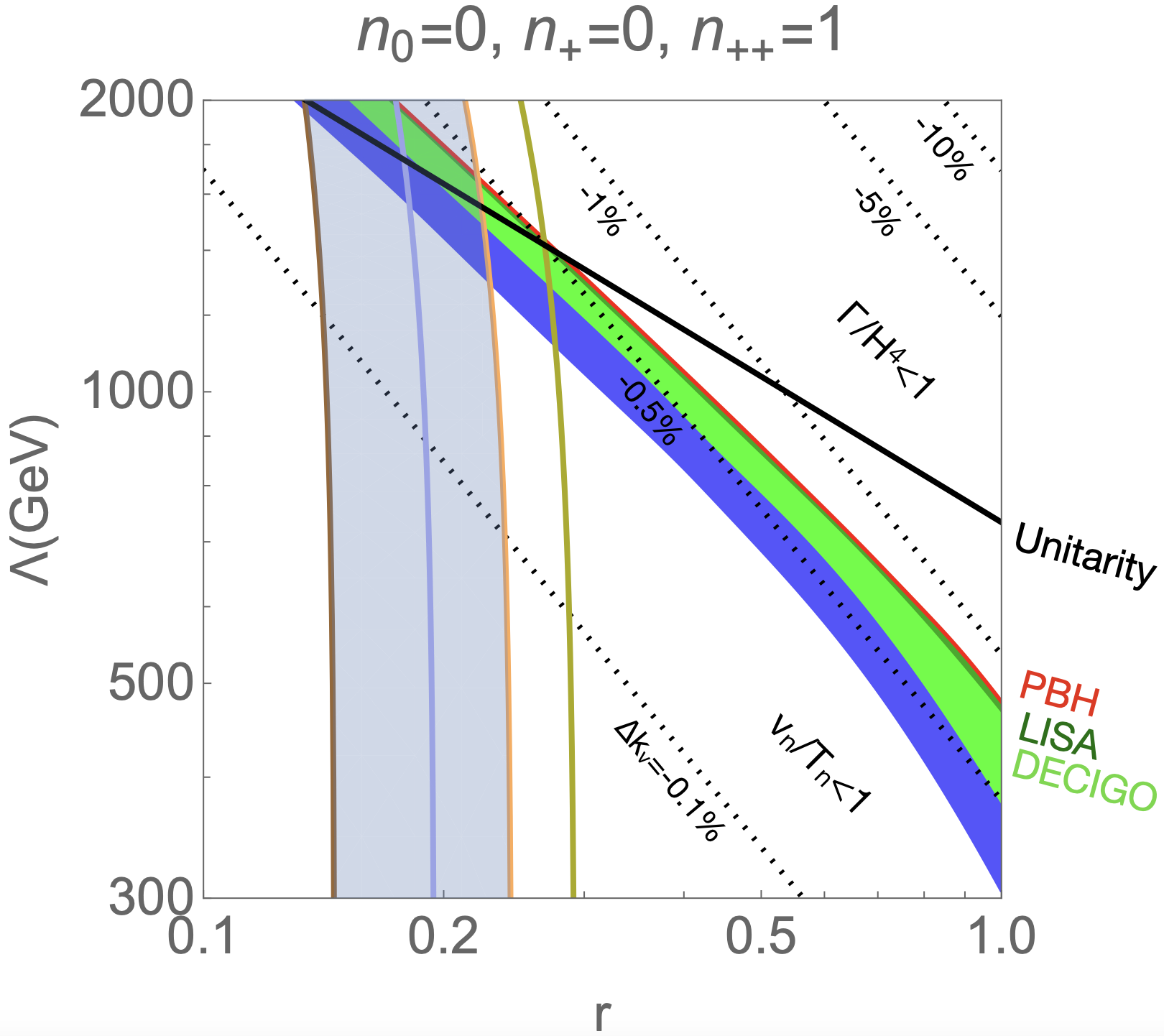}
\includegraphics[width=0.32\textwidth]{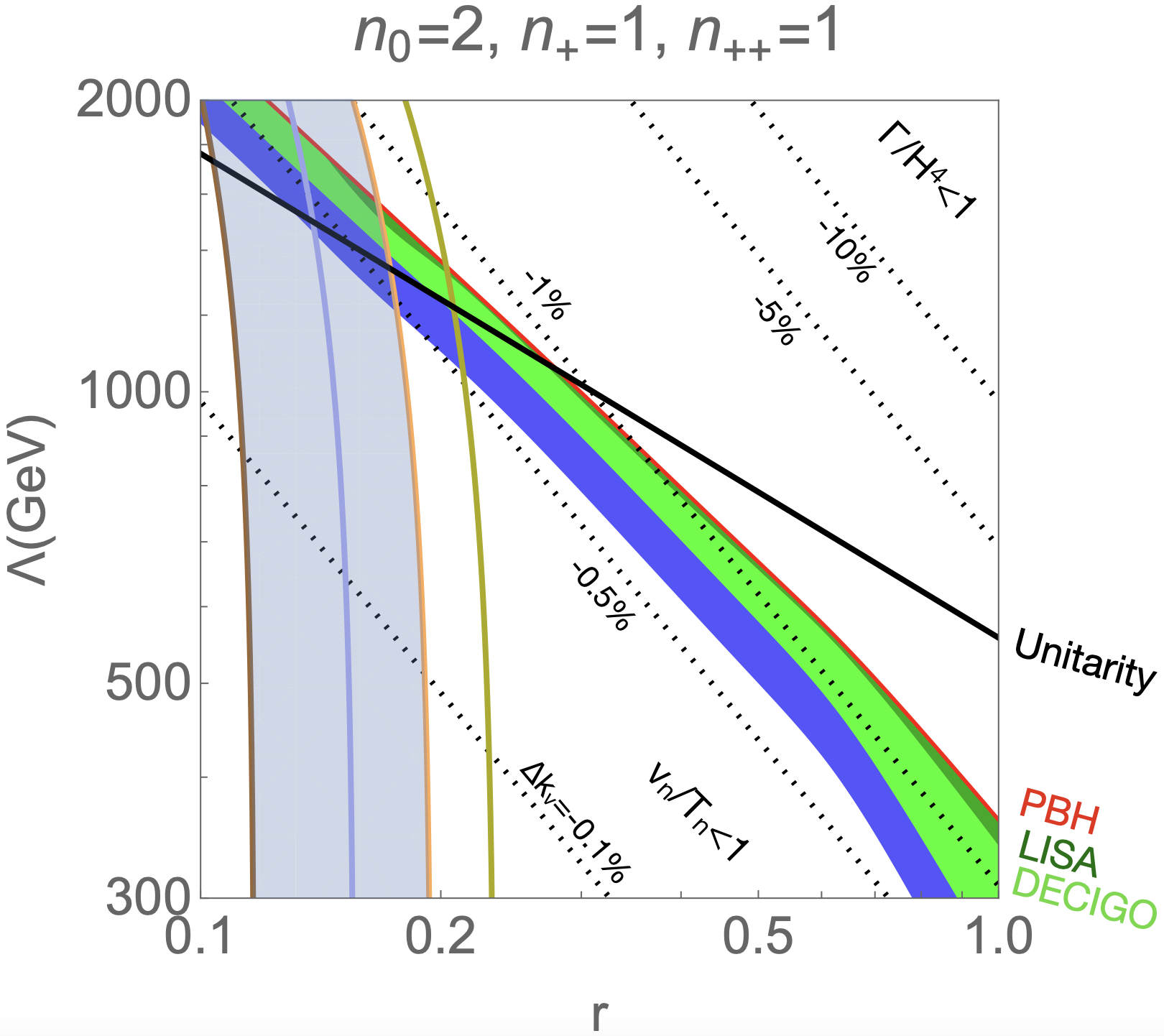}
\includegraphics[width=0.32\textwidth]{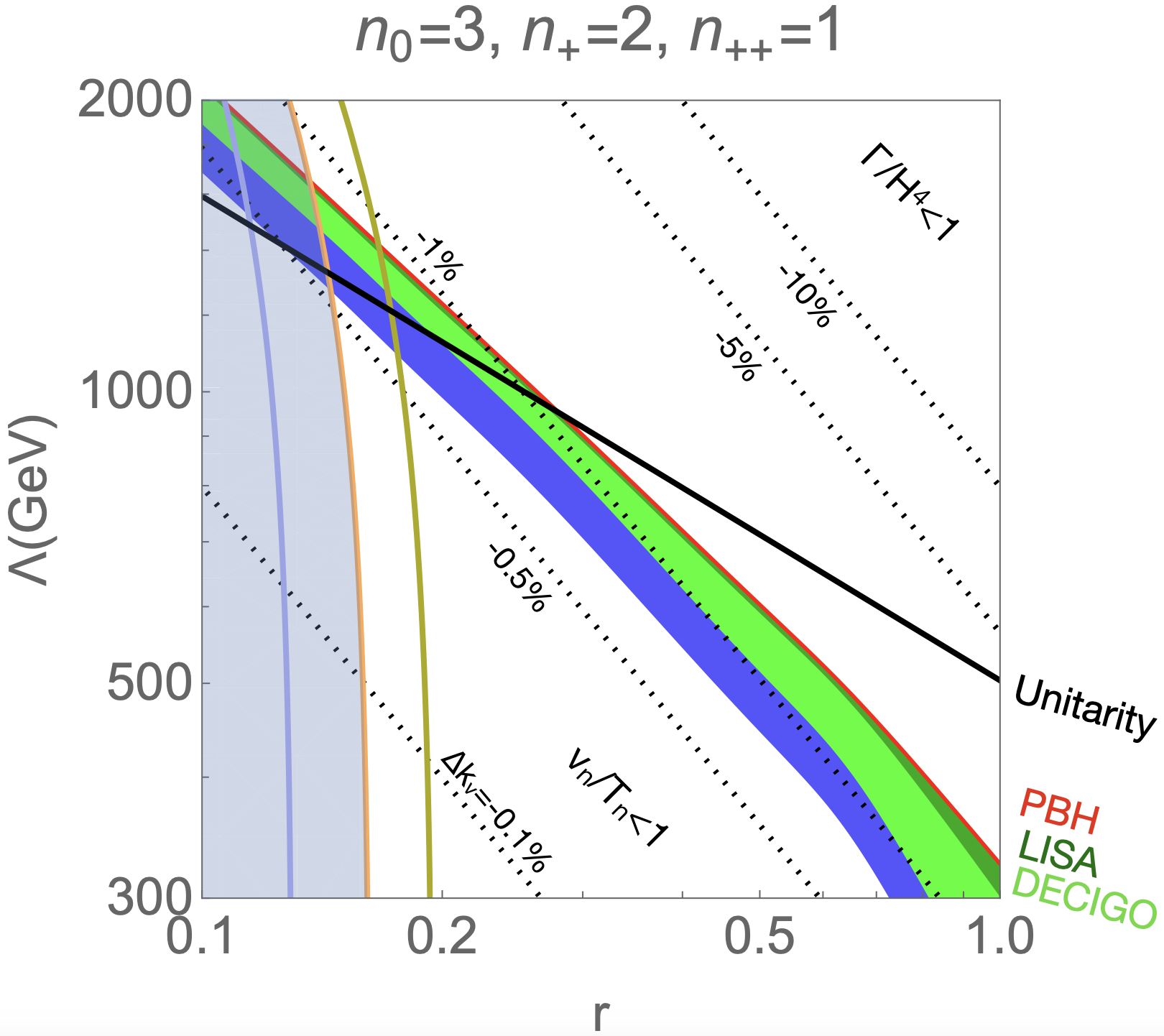}\\
\vspace{0.2cm}
\includegraphics[width=0.75\textwidth]{lh.png}
\caption{
Parameter region explored by $h \gamma \gamma$, $hVV$ and GW observations. 
The parameters $r$ and $\Lambda$ correspond to the non-decouplingness and the mass scale of new particles, respectively.
The black dotted lines are the contours of $hVV$ coupling. 
The value of $\Delta \kappa_{V}$ is defined by $\Delta \kappa_{V} \equiv \kappa_{V} - 1$. 
The definitions of other lines and regions are the same in Fig.~\ref{fig:kg_hhh}. 
\label{fig:kg_hVV}
}
\end{figure}

In Fig.~\ref{fig:kg_hVV}, contours of the $hVV$ coupling are shown instead of the $hhh$ coupling contours in the same setup as Fig.~\ref{fig:kg_hhh}. 
We note that the $hVV$ coupling may be measured with $0.5\,\%$ accuracy by combining the HL-LHC with the ILC with $\sqrt{s} = 250\,{\rm GeV}$ at $68\,\%$ confidence level~\cite{Bambade:2019fyw}. 
Therefore, the parameter region satisfying $\Lambda < 1\,{\rm TeV}$ and Eq.~\eqref{eq:vnTn} can be explored by future precise $hVV$ coupling measurements if $n_{0} \geq 2$. 

Before closing this section, we comment on our analyses. 
In our analyses, it has been assumed that all new particles have the same mass and non-decouplingness. 
In general, a hierarchy between new particle masses can exist.
To discuss the impact of the hierarchy, we consider the case with $(n_{0},\, n_{+}, \, n_{++}) =(2, \, 1, \, 0)$. 
Although this setup is the same as that in the top-right panel of Fig.\,\ref{fig:kg_hhh}, we here assume that the charged particle mass is heavier than the neutral particle mass. 
We also assume that the non-decouplingness of the charged particle is larger than that of the neutral particles.
In Fig.\,\ref{fig:kg_r}, it is shown that charged particles with larger $\Lambda$ and $r$ make the $h \gamma \gamma$ coupling large. 
Therefore, we can expect that the contour for the $h \gamma \gamma$ coupling in the right panel of Fig.~\ref{fig:kg_hhh} is shifted to the left. 
On the other hand, the parameter region satisfying the condition~\eqref{eq:vnTn} is shifted downward. 
Thus, we can expect that the bounds on the $hhh$ coupling and $\Lambda$ we obtained are not drastically changed. 
We have numerically confirmed that our predictions do not much change even if non-decouplingness and mass take slightly different values for each particle.

\begin{table}[t]
\begin{tabular}{|c|c|c|c|c|c|c|}
\hline 
~$(n_{0},\, n_{+}, \, n_{++})$~  & $(0, \, 1, \, 0)$  & $(1, \, 1, \, 0)$ & $(2, \, 1, \, 0)$ & $(0, \, 0, \, 1)$ & $(2, \, 1, \, 1)$ & $(3, \, 2, \, 1)$
\\ \hline 
~Predictions on $\Delta \kappa_{\gamma \gamma}$~ & ~$ - 0.88\, \%$~ &  ~$- 0.97\, \%$~ &  ~$- 0.95\, \%$~  & ~$- 2.71\, \%$~ & ~$- 3.73\, \%$~ & ~$- 4.53\, \%$~
\\ \hline 
\end{tabular}
\caption{
Predictions on the $h \gamma \gamma$ coupling. 
If the measured $h \gamma \gamma$ coupling is larger than these predictions, we can expect that the EWPT can be strongly first-order. \label{table:hgg}}
\end{table}

%%%%%%%%%%%%%%%%%%%%%%%%%%%%%%%%%%%%%%%%%%%%

\section{Discussions and Conclusions \label{sec:conclusion}}

In this paper, we have analyzed the parameter regions in the naHEFT that satisfy the sphaleron decoupling condition in Eq.\,\eqref{eq:vnTn} and the completion condition of the phase transition.
We have investigated the relation between the deviations in the Higgs boson couplings and the dynamics of the EWPT. 
We have employed the naHEFT which can describe the non-decoupling effects in Higgs boson couplings such as $h \gamma \gamma$, $h Z \gamma$, $hVV$ and $hhh$ couplings. 
The expressions of the scaling factors in the naHEFT have been derived. 
We emphasize that the scaling factors in Eq.\,\eqref{eq:scaling_factors} are the general description of Higgs boson couplings in new physics models whose form factors can be described in Eqs.\,\eqref{eq:formKF} and \eqref{eq:FWB} by integrating out new particles.
Although it has been assumed that new particles integrated out have the same non-decoupling property in this paper, our EFT framework can be applicable to models without this assumption.

It has been confirmed that in new physics models with additional charged scalar fields future precision measurements of the $h \gamma \gamma$ coupling can give a useful prediction on the $hhh$ coupling to realize the strongly first-order EWPT.
For instance, in the case with $(n_{0},\, n_{+}, \, n_{++}) =(2, \, 1, \, 0)$, we can expect that the EWPT in the model is strongly first-order if $\Delta \kappa_{\gamma \gamma} < -3\,\%$ and $\Delta \kappa_{3} > 44\, \%$ are confirmed at future collider experiments.
Our results summarized in Tables~\ref{table:hhh} and \ref{table:hgg} play important roles in testing whether the EWPT is strongly first-order or not.
We have also obtained the upper bound on $\Lambda$ by combining precise $h \gamma \gamma$ coupling measurements and the completion condition for the phase transition. 
In the case with $(n_{0},\, n_{+}, \, n_{++}) =(0, \, 1, \, 0)$, it is concluded that a new charged particle satisfying $\Lambda < 946\,{\rm GeV}$ and $r>0.457$ should exist if we confirm $\Delta \kappa_{\gamma \gamma} < -3\,\%$.
We have also confirmed that the $hVV$ coupling measurements are also important to test the strongly first-order EWPT. 
In Fig.\,\ref{fig:kg_hVV}, it has been shown that in the case with $n_{0} >2$ the parameter region satisfying the condition\,\eqref{eq:vnTn} can be explored by precision measurements of $hVV$ couplings at future colliders such as the HL-LHC and ILC.
In addition, the GWs and PBHs produced by the first-order EWPT have been also taken into account. 
In Figs.\,\ref{fig:kg_hhh} and \ref{fig:kg_hVV}, the parameter regions explored by GW and PBH observations have been shown. 

We note that two-loop corrections to $h\gamma \gamma$ and $hhh$ couplings have been evaluated in several renormalizable new physics models~\cite{Braathen:2019pxr,Braathen:2019zoh,Braathen:2020vwo,Degrassi:2023eii,Aiko:2023nqj}. 
It has been confirmed that the two-loop corrections can give a contribution of about 10\% of the one-loop correction. 
In order to describe such higher-order loop corrections, the extension of the current naHEFT is required.

In summary, we have shown that future precision measurements of various Higgs boson couplings including $h \gamma \gamma$ are important to test the strongly first-order EWPT.
We have confirmed that the nature of EWPT and new physics models can be explored by the combination of precision measurements of the Higgs boson couplings at future collider experiments, GW observations at future space-based interferometers and searches for the PBHs.

%%%%%%%%%%%%%%%%%%%%%%%%%%%%%%%%%%%%%%%%%%%%

\begin{acknowledgments}

The work of S. K. was supported in part by the JSPS KAKENHI Grant No. 20H00160 and No. 23K17691.
M. T. was supported by the Iwanami Fujukai Foundation. 

\end{acknowledgments}

\bibliographystyle{apsrev4-2}
\bibliography{FKT_references}

\end{document}